\documentstyle[12pt,psfig,epsfig]{article}
\catcode`\@=11
\long\def\@makefntext#1{
\protect\noindent \hbox to 3.2pt {\hskip-.9pt  
$^{{\ninerm\@thefnmark}}$\hfil}#1\hfill}		

\def\@makefnmark{\hbox to 0pt{$^{\@thefnmark}$\hss}}  
	
\def\ps@myheadings{\let\@mkboth\@gobbletwo
\def\@oddhead{\hbox{}
\rightmark\hfil\ninerm\thepage}   
\def\@oddfoot{}\def\@evenhead{\ninerm\thepage\hfil
\leftmark\hbox{}}\def\@evenfoot{}
\def\sectionmark##1{}\def\subsectionmark##1{}}
\renewcommand{\thefootnote}{\fnsymbol{footnote}}
\newcounter{sectionc}\newcounter{subsectionc}\newcounter{subsubsectionc}
\renewcommand{\section}[1] {\vspace*{0.6cm}\addtocounter{sectionc}{1} 
\setcounter{subsectionc}{0}\setcounter{subsubsectionc}{0}\noindent 
	{\normalsize\bf\thesectionc. #1}\par\vspace*{0.4cm}}
\renewcommand{\subsection}[1] {\vspace*{0.6cm}\addtocounter{subsectionc}{1} 
	\setcounter{subsubsectionc}{0}\noindent 
	{\normalsize\it\thesectionc.\thesubsectionc. #1}\par\vspace*{0.4cm}}
\renewcommand{\subsubsection}[1]
{\vspace*{0.6cm}\addtocounter{subsubsectionc}{1}
	\noindent {\normalsize\rm\thesectionc.\thesubsectionc.\thesubsubsectionc. 
	#1}\par\vspace*{0.4cm}}

\newcounter{appendixc}
\newcounter{subappendixc}[appendixc]
\newcounter{subsubappendixc}[subappendixc]

\renewcommand{\appendix}[1] {\vspace*{0.6cm}
        \refstepcounter{appendixc}
        \setcounter{figure}{0}
        \setcounter{table}{0}
        \setcounter{equation}{0}
        \renewcommand{\thefigure}{\Alph{appendixc}.\arabic{figure}}
        \renewcommand{\thetable}{\Alph{appendixc}.\arabic{table}}
        \renewcommand{\theappendixc}{\Alph{appendixc}}
        \renewcommand{\theequation}{\Alph{appendixc}.\arabic{equation}}
        \noindent{\bf Appendix \theappendixc #1}\par\vspace*{0.4cm}}

\def\abstracts#1{{
	\centering{\begin{minipage}{12.2truecm}\footnotesize\baselineskip=12pt\noindent
	\centerline{\footnotesize ABSTRACT}\vspace*{0.3cm}
	\parindent=0pt #1
	\end{minipage}}\par}} 

\renewenvironment{thebibliography}[1]
	{\begin{list}{\arabic{enumi}.}
	{\usecounter{enumi}\setlength{\parsep}{0pt}
\setlength{\leftmargin 1.25cm}{\rightmargin 0pt}
	 \setlength{\itemsep}{0pt} \settowidth
	{\labelwidth}{#1.}\sloppy}}{\end{list}}
\newcounter{itemlistc}
\newcounter{romanlistc}
\newcounter{alphlistc}
\newcounter{arabiclistc}

\newcommand{\fcaption}[1]{
        \refstepcounter{figure}
        \setbox\@tempboxa = \hbox{\footnotesize Fig.~\thefigure. #1}
        \ifdim \wd\@tempboxa > 6in
           {\begin{center}
        \parbox{6in}{\footnotesize\baselineskip=12pt Fig.~\thefigure. #1}
            \end{center}}
        \else
             {\begin{center}
             {\footnotesize Fig.~\thefigure. #1}
              \end{center}}
        \fi}

\newcommand{\tcaption}[1]{
        \refstepcounter{table}
        \setbox\@tempboxa = \hbox{\footnotesize Table~\thetable. #1}
        \ifdim \wd\@tempboxa > 6in
           {\begin{center}
        \parbox{6in}{\footnotesize\baselineskip=12pt Table~\thetable. #1}
            \end{center}}
        \else
             {\begin{center}
             {\footnotesize Table~\thetable. #1}
              \end{center}}
        \fi}
\def\@citex[#1]#2{\if@filesw\immediate\write\@auxout
	{\string\citation{#2}}\fi
\def\@citea{}\@cite{\@for\@citeb:=#2\do
	{\@citea\def\@citea{,}\@ifundefined
	{b@\@citeb}{{\bf ?}\@warning
	{Citation `\@citeb' on page \thepage \space undefined}}
	{\csname b@\@citeb\endcsname}}}{#1}}

\newif\if@cghi
\def\cite{\@cghitrue\@ifnextchar [{\@tempswatrue
	\@citex}{\@tempswafalse\@citex[]}}
\def\citelow{\@cghifalse\@ifnextchar [{\@tempswatrue
	\@citex}{\@tempswafalse\@citex[]}}
\def\@cite#1#2{{$\null^{#1}$\if@tempswa\typeout
	{IJCGA warning: optional citation argument 
	ignored: `#2'} \fi}}

 1
 1
 1

\font\ninerm=cmr9

\def\ifmath#1{\relax\ifmmode #1\else $#1$\fi}
\def\half{\ifmath{{\textstyle{1 \over 2}}}}

\def\3quarter{{\textstyle{3 \over 4}}}

\def\lf{\leaders\hbox to 1em{\hss.\hss}\hfill}
\def\e6{$E(6)$}
\def\10{$SO(10)$}
\def\21{$SU(2) \otimes U(1) $}
\def\lr{$SU(2)_L \otimes SU(2)_R \otimes U(1)$}
\def\422{$SU(4) \otimes SU(2) \otimes SU(2)$}
\def\321{$SU(3) \otimes SU(2) \otimes U(1)$}
\def\ne{\hbox{$\nu_e$ }}
\def\nm{\hbox{$\nu_\mu$ }}
\def\nt{\hbox{$\nu_\tau$ }}
\def\ns{\hbox{$\nu_{s}$ }}

\def\O{\hbox{$\cal O$ }}

\def\mnt{\hbox{$m_{\nu_\tau}$ }}

\def\neus{\hbox{neutrinos }}

\def\neu{\hbox{neutrino }}

\def\eq#1{{eq. (\ref{#1})}}

\def\fig#1{{Fig. (\ref{#1})}}

\let\vev\VEV

\def\lsim{\raise0.3ex\hbox{$\;<$\kern-0.75em\raise-1.1ex\hbox{$\sim\;$}}}
\def\gsim{\raise0.3ex\hbox{$\;>$\kern-0.75em\raise-1.1ex\hbox{$\sim\;$}}}
\def\half{{1\over 2}}
\def\beq{\begin{equation}}
\def\eeq{\end{equation}}
\def\bef{\begin{figure}}
\def\eef{\end{figure}}
\def\bet{\begin{table}}
\def\eet{\end{table}}
\def\bea{\begin{eqnarray}}
\def\ba{\begin{array}}
\def\ea{\end{array}}
\def\bi{\begin{itemize}}
\def\ei{\end{itemize}}
\def\ben{\begin{enumerate}}
\def\een{\end{enumerate}}

\def\eea{\end{eqnarray}}
\def\apj#1#2#3{          {\it Astrophys. J. }{\bf #1} (19#2) #3}

\def\aa#1#2#3{          {\it Astron. \& Astrophys.  }{\bf #1} (19#2) #3}

\def\ib#1#2#3{           {\it ibid. }{\bf #1} (19#2) #3}
\def\nat#1#2#3{          {\it Nature }{\bf #1} (19#2) #3}
\def\nps#1#2#3{        {\it Nucl. Phys. B (Proc. Suppl.) }{\bf #1} (19#2) #3} 
\def\np#1#2#3{           {\it Nucl. Phys. }{\bf #1} (19#2) #3}
\def\pl#1#2#3{           {\it Phys. Lett. }{\bf #1} (19#2) #3}
\def\pr#1#2#3{           {\it Phys. Rev. }{\bf #1} (19#2) #3}
\def\prep#1#2#3{         {\it Phys. Rep. }{\bf #1} (19#2) #3}
\def\prl#1#2#3{          {\it Phys. Rev. Lett. }{\bf #1} (19#2) #3}

\def\zp#1#2#3{           {\it Zeit. fur Physik }{\bf #1} (19#2) #3}

\def\n.c.#1#2#3{         {\it Nuovo Cim. }{\bf #1} (19#2) #3}
\def\r.n.c.#1#2#3{       {\it Riv. del Nuovo Cim. }{\bf #1} (19#2) #3}
\def\sjnp#1#2#3{         {\it Sov. J. Nucl. Phys. }{\bf #1} (19#2) #3}

\def\mpl#1#2#3{          {\it Mod. Phys. Lett. }{\bf #1} (19#2) #3}

\def\ppnp#1#2#3{           {\it Prog. Part. Nucl. Phys. }{\bf #1} (19#2) #3}

\def\pc{private communication}

\def\ip{in preparation}
\def\bne{\hbox{$\bar\nu_e$ }}  
\def\bnm{\hbox{$\bar\nu_\mu$ }}  
 
\begin{document}
\centerline{\normalsize\bf NEUTRINO MASSES, WHERE DO WE STAND?
\footnote{Proceedings of New Trends in Neutrino Physics,
May 1998, Ringberg Castle, Tegernsee, Germany. Presented 
also as a set of three lectures at the V Gleb Wataghin 
School on High Energy Phenomenology, Campinas, Brazil, July 1998}
}
\baselineskip=22pt
\centerline{\footnotesize JOS\'E W. F. VALLE}
\baselineskip=13pt
\centerline{\footnotesize \it Inst. de F\'{\i}sica Corpuscular 
- C.S.I.C. - Dept. de F\'{\i}sica Te\`orica, Univ. de
Val\`encia}
\baselineskip=12pt
\centerline{\footnotesize \it 46100 Burjassot, Val\`encia, Spain}
\centerline{\footnotesize http://neutrinos.uv.es }

\vspace*{0.9cm} 

\abstracts{I review the status of neutrino physics post-Neutrino~98,
including the implications of solar and atmospheric neutrino data,
which strongly indicate nonzero neutrino masses.  LSND and the
possible role of neutrinos as hot dark matter (HDM) are also
mentioned.  The simplest schemes proposed to reconcile these
requirements invoke a light sterile neutrino in addition to the three
active ones, two of them at the MSW scale and the other two
maximally-mixed neutrinos at the HDM/LSND scale. In the simplest
theory the latter scale arises at one-loop, while the solar and
atmospheric parameters $\Delta {m^2}_\odot$ \& $\Delta {m^2}_{atm}$
appear at the two-loop level. The lightness of the sterile neutrino,
the nearly maximal atmospheric neutrino mixing, and the generation of
$\Delta {m^2}_\odot$ \& $\Delta {m^2}_{atm}$ follow naturally from the
assumed lepton-number symmetry and its breaking.  These two basic
schemes can be distinguished at future solar \& atmospheric neutrino
experiments and have different cosmological implications. }
\normalsize\baselineskip=15pt
\setcounter{footnote}{0}
\renewcommand{\thefootnote}{\alph{footnote}}

\section{Introduction}
\vskip .1cm

Neutrinos are the only fermions which the Standard Model (SM) predicts
to be massless.  This ansatz was justified due to the apparently
masslessness of neutrinos in most experiments. However, the situation
has changed due to the important impact of underground experiments,
since the pioneer geochemical experiments of Davis and collaborators,
to the more recent Gallex, Sage, Kamiokande and SuperKamiokande
experiments \cite{solarexp,atmexp,superkatm98,sk300,sk504}. Altogether
they provide solid evidence for the solar and the atmospheric neutrino
problems, two milestones in the search for physics beyond the SM. Of
particular importance has been the recent confirmation by the
SuperKamiokande collaboration \cite{superkatm98} of the atmospheric
neutrino zenith-angle-dependent deficit, which has marked a turning
point in our understanding of neutrinos, providing a strong evidence
for \nm conversions.  In addition to the neutrino data from
underground experiments there is also some possible indication for
neutrino oscillations from the LSND experiment~\cite{LSND}. To this we
may add the possible r\^ole of neutrinos in the dark matter problem
and structure formation \cite{cobe,cobe2,iras}.  If one boldly insists
in including also the last two requirements, together with the data on
solar and atmospheric neutrinos, then we have {\sl three mass scales}
involved in neutrino oscillations. The simplest way to reconcile these
requirements invokes the existence of a light sterile neutrino
\cite{ptv92,pv93,cm93}. The prototype models proposed in
\cite{ptv92,pv93} enlarge the \21 Higgs sector in such a way that
neutrinos acquire mass radiatively, without unification nor
seesaw. Out of the four neutrinos, two of them lie at the MSW scale
and the other two maximally-mixed neutrinos are at the HDM/LSND
scale. The latter scale arises at one-loop, while the solar and
atmospheric scales come in at the two-loop level. The lightness of the
sterile neutrino, the nearly maximal atmospheric neutrino mixing, and
the generation of the solar and atmospheric neutrino scales all result
naturally from the assumed lepton-number symmetry and its breaking.
Either \ne- \nt conversions explain the solar data with
\nm- \ns oscillations accounting for the atmospheric deficit
\cite{ptv92}, or else the r\^oles of \nt and \ns are reversed
~\cite{pv93}. These two basic schemes have distinct implications at
future solar \& atmospheric neutrino experiments, as well as
cosmology.

\section{Theories of Neutrino Mass}
\vskip .1cm

One of the most unpleasant features of the SM is that the masslessness
of neutrinos is not dictated by an underlying {\sl principle}, such as
that of gauge invariance in the case of the photon: the SM simply
postulates that neutrinos are massless by choosing a restricted
multiplet content.  {\sl Why are neutrinos so special when compared
with the other fundamental fermions}?  If massive, neutrinos would
present another puzzle: {\sl Why are their masses so small compared to
those of the charged fermions}? The fact that neutrinos are the only
electrically neutral elementary fermions may hold the key to the
answer, namely neutrinos could be Majorana fermions, the most
fundamental kind of fermion. In this case the suppression of their
mass could be associated to the breaking of lepton number symmetry at
a very large energy scale within a {\sl unification approach}, which
can be implemented in many extensions of the SM. Alternatively,
neutrino masses could arise from garden-variety {\sl weak-scale
physics} characterized by a scale $\vev{\sigma} = \O(m_Z)$ where
$\vev{\sigma}$ denotes a \21 singlet vacuum expectation value which
owes its smallness to the symmetry enhancement which would result if
$\vev{\sigma}$ and $m_\nu \to 0$.

One should realize however that, although the physics of neutrinos can
be rather different in various gauge theories of neutrino mass, there
is hardly any predictive power on masses and mixings, this is one of
the aspects of the so-called flavour problem which is probably the
toughest open problem in physics.

\subsection{Unification Approach}
\vskip .1cm

An attractive possibility is to ascribe the origin of parity violation
in the weak interaction to the spontaneous breaking of B-L symmetry
in the context of left-right symmetric extensions such as the \lr
\cite{LR}, \422 \cite{PS} or \10 gauge groups \cite{GRS}. In this case
the masses of the light neutrinos are obtained by diagonalizing the
following mass matrix in the basis $\nu,\nu^c$
\begin{equation}
\left[\matrix{
 M_L & D \cr
 D^T & M_R }\right] 
\label{SS} 
\end{equation} 
where $D$ is the standard \21 breaking Dirac mass term and $M_R =
M_R^T$ is the isosinglet Majorana mass that may arise from a 126
vacuum expectation value (vev) in \10. The magnitude of the $M_L
\nu\nu$ term \cite{2227} is also suppressed by the left-right breaking
scale, $M_L \propto 1/M_R$ \cite{LR}.

In the seesaw approximation, one finds 
\beq 
M_{\nu \: eff} = M_L - D M_R^{-1} D^T
\:.
\label{SEESAW} 
\eeq 
As a result one is able to explain naturally the relative smallness of
\neu masses since $m_\nu \propto 1/M_R$.  Although $M_R$ is expected
to be large, its magnitude heavily depends on the model and it may
have different possible structures in flavour space (so-called
textures) \cite{Smirnov}.  As a result it is hard to make firm
predictions for the corresponding light neutrino masses and mixings
that are generated through the seesaw mechanism. In fact this freedom
has been exploited in model building in order to account for an almost
degenerate seesaw-induced neutrino mass spectrum \cite{DEG}.

One virtue of the unification approach is that it may allow one to
gain a deeper insight into the flavour problem. There have been
interesting attempts at formulating supersymmetric unified schemes
with flavour symmetries and texture zeros in the Yukawa couplings.  In
this context a challenge is to obtain the large lepton mixing now
indicated by the atmospheric neutrino data.

\subsection{Weak-Scale Approach}
\vskip .1cm

Although very attractive, the unification approach is by no means the
only way to generate neutrino masses. There are many schemes which do
not require any large mass scale.  The extra particles employed to
generate the neutrino masses have masses $\O(m_Z)$ accessible to
present experiments. There is a variety of such mechanisms, in which
neutrinos acquire mass either at the tree level or radiatively.  Let
us look at some.

\subsubsection{Tree Level}
\vskip .1cm

For example, it is possible to extend the lepton sector of the \21
theory by adding a set of $two$ 2-component isosinglet neutral
fermions, denoted ${\nu^c}_i$ and $S_i$, $i=e,~\mu$ or $\tau$ in each
generation. In this case one can consider the mass matrix (in the
basis $\nu, \nu^c, S$) \cite{CON}
\begin{equation}
\left[\matrix{
  0 & D & 0 \cr
  D^T & 0 & M \cr
  0 & M^T & \mu }\right] 
\label{MATmu} 
\end{equation} 
The Majorana masses for the neutrinos are determined from
\beq
M_L = D M^{-1} \mu {M^T}^{-1} D^T
\label{33}
\eeq 
In the limit $\mu \to 0$ the exact lepton number symmetry is recovered
and will keep neutrinos strictly massless to all orders in
perturbation theory, as in the SM. The corresponding texture of the
mass matrix has been suggested in various theoretical models
\cite{WYLER}, such as superstring inspired models~\cite{SST}. In the
latter the zeros arise due to the lack of Higgs fields to provide the
usual Majorana mass terms.
The smallness of neutrino mass then follows from the smallness of
$\mu$. The scale characterizing $M$, unlike $M_R$ in the seesaw
scheme, can be low. As a result, in contrast to the heavy neutral
leptons of the seesaw scheme, those of the present model can be light
enough as to be produced at high energy colliders such as LEP
\cite{CERN} or at a future Linear Collider. The smallness of $\mu$ is
in turn natural, in t'Hooft's sense, as the symmetry increases when
$\mu \to 0$, i.e.  total lepton number is restored.  This scheme is a
good alternative to the smallness of neutrino mass, as it bypasses
the need for a large mass scale, present in the seesaw unification
approach. One can show that, since the matrices $D$ and $M$ are not
simultaneously diagonal, the leptonic charged current exhibits a
non-trivial structure that cannot be rotated away, even if we set $\mu
\equiv 0$.  The phenomenological implication of this, otherwise
innocuous twist on the SM, is that there is neutrino mixing despite
the fact that light neutrinos are strictly massless.  It follows that
flavour and CP are violated in the leptonic currents, despite the
masslessness of neutrinos. The loop-induced lepton flavour and CP
non-conservation effects, such as $\mu \to e + \gamma$~\cite{BER,3E},
or CP asymmetries in lepton-flavour-violating processes such as $Z \to
e \bar{\tau}$ or $Z \to \tau \bar{e}$~\cite{CP} are precisely
calculable. The resulting rates may be of experimental interest
\cite{ETAU,TTTAU,cernlfv}, since they are not constrained by the
bounds on neutrino mass, only by those on universality, which are
relatively poor.  In short, this is a conceptually simple and
phenomenologically rich scheme.

Another remarkable implication of this model is a new type of resonant
neutrino conversion mechanism \cite{massless0}, which was the first
resonant mechanism to be proposed after the MSW effect \cite{MSW}, in
an unsuccessful attempt to bypass the need for neutrino mass in the
resolution of the solar neutrino problem. According to the mechanism,
massless neutrinos and anti-neutrinos may undergo resonant flavour
conversion, under certain conditions. Though these do not occur in the
Sun, they can be realized in the chemical environment of supernovae
\cite{massless}. Recently it has been pointed out how they may provide
an elegant approach for explaining the observed velocity of pulsars
\cite{pulsars}.
 
\subsubsection{Radiative Level}
\vskip .1cm

\begin{figure}[t]
\centerline{\protect\hbox{\psfig{file=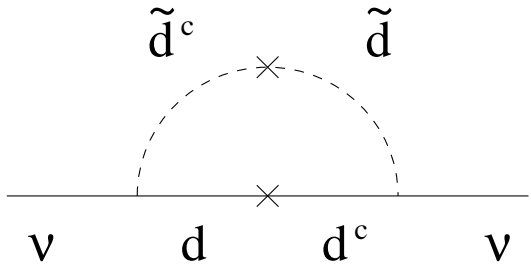,height=4cm,width=7cm}}}
\vglue -0.6cm
\caption{One-loop-induced Neutrino Mass. }
\label{mnrad}
\end{figure}

There is also a large variety of {\sl radiative} models, where the \21
multiplet content is extended in order to generate neutrino
masses. The prototype one-loop scheme is the one proposed by Zee
\cite{zee}. Supersymmetry with explicitly broken R-parity also
provides an alternative one-loop mechanism to generate neutrino mass.
These arise, for example, from scalar quark or scalar lepton
contributions, as shown in \fig{mnrad}

\begin{figure}
\centerline{\protect\hbox{\psfig{file=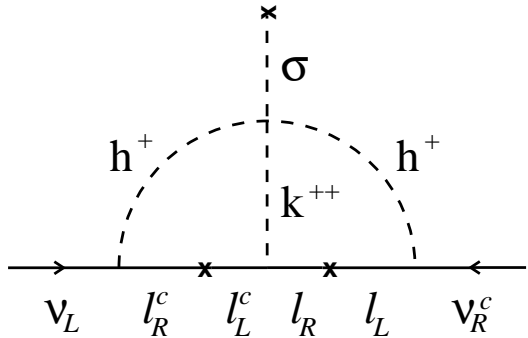,height=4.5cm,width=7cm}}}
\caption{Two-loop-induced Neutrino Mass }
\label{2loop}
\end{figure}

A two-loop scheme to induce neutrino mass was suggested by Babu
\cite{Babu88}. The relevant diagram is shown in \fig{2loop}
\footnote{Note here that I have used the slight variant of the Babu
model suggested in ref. \cite{ewbaryo}, which incorporates the idea of
spontaneous, rather than explicit lepton number violation}.

In the above examples active neutrinos acquire radiative mass. One can
also employ the radiative approach to construct models including
sterile neutrinos, such as those in ref.~\cite{ptv92,pv93}. In this
case some new Feynman diagram topologies are encountered.

\subsection{A Hybrid Approach}
\vskip .1cm

I now describe an interesting mechanism of neutrino mass generation
that combines seesaw and radiative mechanisms. It invokes
supersymmetry with broken R-parity, as the origin of neutrino mass and
mixings \cite{epsrad}. The simplest model is a unified minimal
supergravity model with universal soft breaking parameters (MSUGRA)
and bilinear breaking of R--parity~\cite{epsrad,RPothers}. Contrary to
a popular misconception, the bilinear violation of R--parity implied
by the $\epsilon_3$ term in the superpotential is physical, and can
not be rotated away~\cite{BRpVtalks}. It leads also by a minimization
condition, to a non-zero sneutrino vev, $v_3$.  It is
well-known~\cite{rossarca} that in such models of broken R--parity the
tau neutrino $\nu_{\tau}$ acquires a mass, due to the mixing between
neutrinos and neutralinos. It comes from the matrix
\begin{equation}
\left[\matrix{
M_1 & 0  & -\half g'v_d & \half g'v_u & -\half g'v_3 \cr
0   & M_2 & \half g v_d & -\half g v_u & \half g v_3 \cr
-\half g'v_d & \half g v_d & 0 & -\mu & 0 \cr
\half g'v_u & -\half g v_u & -\mu & 0 & \epsilon_3 \cr
-\half g'v_3 & \half g v_3 & 0 & \epsilon_3 & 0 
}\right]
\label{eq:NeutMassMat}
\end{equation}
where the first two rows are gauginos, the next two Higgsinos, and the
last one denotes the tau neutrino. The $v_u$ and $v_d$ are the
standard vevs, $g's$ are gauge couplings and $M_{1,2}$ are the gaugino
mass parameters. Since the $\epsilon_3$ and the $v_3$ are related, the
simplest (one-generation) version of this model contains only one
extra free parameter in addition to those of the MSUGRA model.  The
universal soft supersymmetry-breaking parameters at the unification
scale $m_X$ are evolved via renormalization group equations down to
the weak scale $\O(m_Z)$.  This induces an effective non-universality
of the soft terms {\sl at the weak scale} which in turn implies a
non-zero sneutrino vev $v'_3$ given as
\begin{equation}
v'_3 \approx \frac{\epsilon_3 \mu} {{m_Z}^4}
\left(v'_d \Delta M^2 + \mu'v_u \Delta B \right)
\label{App_v3p}
\end{equation}
where the primed quantities refer to a basis in which we eliminate the
$\epsilon_3$ term from the superpotential (but reintroduce it, of
course, in other sectors of the theory).  

The scalar soft masses and bilinear mass parameters obey $\Delta
M^2=0$ and $\Delta B=0$ at $m_X$. However at the weak scale they are
calculable from radiative corrections as
\begin{eqnarray}
\Delta M^2 & \approx & {{3h_b^2} \over{8\pi^2}} m_{Z}^2
\ln{{M_{GUT}}\over{m_Z}}
\end{eqnarray}
Note that \eq{App_v3p} implies that the R--parity-violating effects
induced by $v'_3$ are {\sl calculable} in terms of the primordial
R--parity-violating parameter $\epsilon_3$. It is clear that the
universality of the soft terms plays a crucial r\^ole in the
calculability of the $v'_3$ and hence of the resulting neutrino mass
\cite{epsrad}. Thus \eq{eq:NeutMassMat} represents a new kind of
see-saw scheme in which the $M_R$ of \eq{SS} is the neutralino mass,
while the r\^ole of the Dirac entry $D$ is played by the $v'_3$, which
is induced radiatively as the parameters evolve from $m_X$ to the weak
scale. Thus we have a {\sl hybrid} see-saw mechanism, with naturally
suppressed Majorana $\nu_{\tau}$ mass induced by the mixing between
the weak eigenstate tau neutrino and the {\sl zino}.

\begin{figure}[t]
\centerline{\protect\hbox{\psfig{file=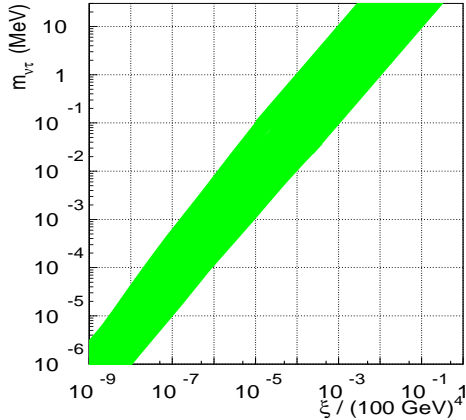,height=6cm,width=7cm}}}
\caption{Tau neutrino mass versus 
$\xi \equiv(\epsilon_3v_d+\mu v_3)^2$, from ref.~\protect\cite{epsrad} }
\label{mnt_xi_ev}
\end{figure}

Let me now turn to estimate the expected \nt mass. For this purpose
let me first determine the tau neutrino mass in the most general
supersymmetric model with bilinear breaking of R-parity, {\sl without
imposing soft universality}.  The \nt mass depends quadratically on an
effective parameter $\xi$ defined as $\xi \equiv (\epsilon_3 v_d + \mu
v_3)^2 \propto {v'_3}^2$ characterizing the violation of R--parity.
The expected \mnt values are illustrated in \fig{mnt_xi_ev}. The band
shown in the figure is obtained through a scan over the parameter
space requiring that the supersymmetric particles are not too light.
Let us now compare this with the cosmologically allowed values of the
tau neutrino mass. The cosmological critical density bound \mnt $\lsim
92 \Omega h^2$ eV only holds if neutrinos are stable.  In the present
model (with 3-generations) the \nt can decay into 3 neutrinos, via the
neutral current \cite{2227,774}, or by slepton exchanges. This decay
will reduce the relic \nt abundance to the required level, as long as
\nt is heavier than about 100 KeV or so. On the other hand primordial
Big-Bang nucleosynthesis implies that \nt is lighter than about an MeV
or so \cite{bbnutaustable}.

However, {\sl if one adopts a SUGRA scheme where universality of the
soft supersymmetry breaking terms at $m_X$ is assumed}, then the \nt
mass is theoretically {\sl predicted} in terms of $h_b$ and can be
small in this case due to a natural cancellation between the two terms
in the parameter $\xi$, which follows from the assumed universality of
the softs at $m_X$. One can verify that \mnt may easily lie in the
electron-volt range, in which case \nt could be a component of the hot
dark matter of the Universe.

Notice that \ne and \nm remain massless in this approximation. They
get masses either from scalar loop contributions in \fig{mnrad} or by
mixing with singlets in models with spontaneous breaking of R-parity
\cite{Romao92}.  It is important to notice that even when \mnt is
small, many of the corresponding R-parity violating effects can be
sizeable. An obvious example is the fact that the lightest neutralino
decay will typically decay inside the detector, unlike standard
R-parity-conserving supersymmetry. This leads to a vastly unexplored
plethora of phenomenological possibilities in supersymmetric physics
\cite{desert}.

\vskip .1cm

In conclusion I can say that, other than the seesaw scheme, none of
the above models requires a large mass scale. As a result they lead to
a potentially rich phenomenology, since the extra particles required
have masses at scales that could be accessible to present experiments.
In the simplest versions of these models the neutrino mass arises from
the explicit violation of lepton number. Their phenomenological
potential gets richer if one generalizes the models so as to implement
a spontaneous violation scheme. This brings me to the next section.

\subsection{Weak-scale majoron}
\vskip .1cm

If lepton number (or B-L) is an ungauged symmetry and if it is
arranged to break spontaneously, the generation of neutrino masses
will be accompanied by the existence of a physical Goldstone boson
that we generically call majoron.
Except for the left-right symmetric unification approach, in which B-L
is a gauge symmetry, in all of the above schemes one can implement the
spontaneous violation of lepton number. 
One can also introduce it in an \21 seesaw framework \cite{CMP}, as
originally proposed, but I do not consider this case here, see
ref.~\cite{fae} for a review. Here I will mainly concentrate on
weak-scale physics. In all models I consider the lepton-number breaks
at a scale given by a vacuum expectation value $\vev{\sigma} \sim
m_{weak}$.
Such scale arises as the most natural one since in all of these
models, as already mentioned, we have that the neutrino masses vanish
as the lepton-breaking scale $\vev{\sigma} \to 0$
\cite{JoshipuraValle92}.

It is also clear that in any acceptable model one must arrange for the
majoron to be mainly an \21 singlet, ensuring that it does not affect
the invisible Z decay width, well-measured at LEP. In models where the
majoron has L=2 the neutrino mass is proportional to an insertion of
$\vev{\sigma}$, as indicated in \fig{2loop}.  In the supersymmetric
model with broken R-parity the majoron is mainly a singlet sneutrino,
which has lepton number L=1, so that $m_\nu \propto \vev{\sigma}^2$,
where $\vev{\sigma} \equiv \vev{\widetilde{\nu^c}}$, with
$\widetilde{\nu^c}$ denoting the singlet sneutrino. The presence of
the square, just as in the parameter $\xi$  ~in \fig{mnt_xi_ev},
reflects the fact that the neutrino gets a Majorana mass which has
lepton number L=2.  The sneutrino gets a vev at the effective
supersymmetry breaking scale $ m_{susy} = m_{weak}$.

The weak-scale majorons may have remarkable phenomenological
implications, such as the possibility of invisibly decaying Higgs bosons
\cite{JoshipuraValle92}.  Unfortunately I have no time to discuss it
here (see, for instance \cite{desert}).  

If the majoron acquires a KeV mass (natural in weak-scale models) from
gravitational effects at the Planck scale \cite{ellis} it may play a
r\^ole in cosmology as dark matter~\cite{KEV}.
In what follows I will just focus on two examples of how the
underlying physics of weak-scale majoron models can affect neutrino
cosmology in an important way.

\subsubsection{Heavy neutrinos and the Universe Mass}
\vskip .1cm

Neutrinos of mass less than \O(100 KeV) or so, are cosmologically
stable if they have only SM interactions. Their contribution to the
present density of the universe implies \cite{KT}
\beq 
\label{RHO1}
\sum m_{\nu_i} \lsim 92 \: \Omega_{\nu} h^2 \: eV\:, 
\eeq 
where the sum is over all isodoublet neutrino species with mass less
than \O(1 MeV). The parameter $\Omega_{\nu} h^2 \leq 1$, where $h^2$
measures the uncertainty in the present value of the Hubble parameter,
$0.4 \lsim h \lsim 1$, while $\Omega_{\nu} = \rho_{\nu}/\rho_c$,
measures the fraction of the critical density $\rho_c$ in neutrinos.
For the $\nu_{\mu}$ and $\nu_{\tau}$ this bound is much more stringent
than the laboratory limits.

In weak-scale majoron models the generation of neutrino mass is
accompanied by the existence of a physical majoron, with potentially
fast majoron-emitting decay channels such as \cite{fae,V}
\beq
\label{NUJ}
\nu^\prime  \to \nu + J \:\: .
\eeq
as well as new annihilations to majorons,
\beq
\label{nunuJJ}
\nu^\prime  + \nu^\prime  \to J + J \:\: .
\eeq
These could eliminate relic neutrinos and therefore allow neutrinos of
higher mass, as long as the rates are large enough to allow for an
adequate red-shift of the heavy neutrino decay and/or annihilation
products. While the annihilation involves a diagonal majoron-neutrino
coupling $g$, the decays proceed only via the non-diagonal part of the
coupling, in the physical mass basis.  A careful diagonalization of
both mass matrix and coupling matrix is essential in order to avoid
wild over-estimates of the heavy neutrino decay rates, such as that in
ref.~\cite{CMP}.
The point is that, once the neutrino mass matrix is diagonalized,
there is a danger of simultaneously diagonalizing the majoron
couplings to neutrinos. That would be analogous to the GIM mechanism
present in the SM for the couplings of the Higgs to fermions. Models
that avoid this GIM mechanism in the majoron-neutrino couplings have
been proposed, e.g. in ref.~\cite{V}. Many of them are weak-scale
majoron models \cite{CON,JoshipuraValle92,Romao92}.  A general method
to determine the majoron couplings to neutrinos and hence the neutrino
decay rates in any majoron model was first given in ref. \cite{774}.
For an estimate in the model with spontaneously broken R-parity
\cite{MASIpot3} see ref. \cite{Romao92}.

In short one may say that neutrino lifetimes can be shorter than
required by the cosmological mass bound, for all values of the masses
which are presently allowed by laboratory experiments. 

\subsubsection{Heavy neutrinos and  Cosmological Nucleosynthesis} 
\vskip .1cm

\begin{figure}
\centerline{\protect\hbox{\psfig{file=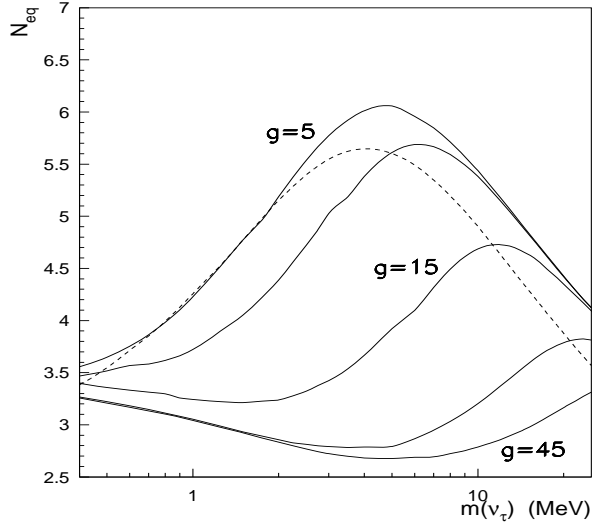,height=7cm,width=8cm}}}
\caption{The dashed line shows the effective number of massless SM
neutrinos equivalent to the heavy \nt ($g=0$). Depending on the value
of $g$ (in units of $10^{-5}$) one can lower $N_\nu$ below the
canonical SM value $N_\nu = 3$ due to the effect of \nt
annihilations. From ref. \protect\cite{DPRV} }
\label{bbneq}
\end{figure}

Similarly, the number of light neutrino species is also restricted by
cosmological Big Bang Nucleosynthesis (BBN). Due to its large mass, an
MeV stable (lifetime longer than $\sim 100$ sec) tau neutrino would be
equivalent to several SM massless neutrino species and would therefore
substantially increase the abundance of primordially produced
elements, such as $^{4}He$ and deuterium
\cite{dmeas,cris.ncris,sarkar}.  This can be converted into
restrictions on the \nt mass. If the bound on the effective number of
massless neutrino species is taken as $N_\nu < 3.4-3.6$, one can rule
out $\nu_\tau$ masses above 0.5 MeV~\cite{bbnutaustable}.  If we take
$N_\nu < 4.5$ \cite{sarkar} the \mnt limit loosens accordingly, as
seen from \fig{bbneq}, and allows a \nt of about an MeV or so.

\begin{figure}
\centerline{
\psfig{file=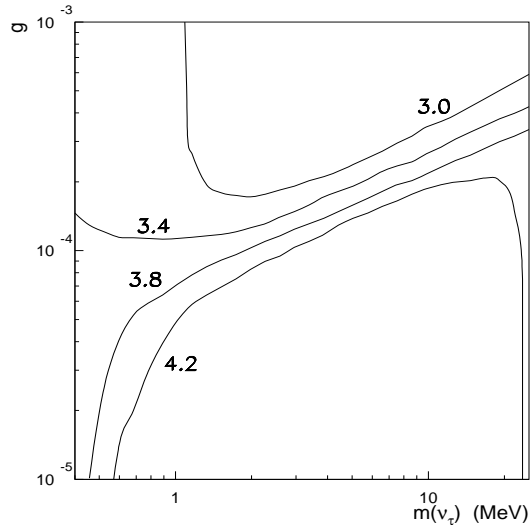,height=7cm,width=7cm}}
\caption{The region above each curve is allowed for the corresponding
$N_\nu^{max}$. From ref. \protect\cite{DPRV} } \vglue -.5cm
\vglue -.5cm
\label{neffmg}
\end{figure}

In the presence of \nt annihilations the BBN \mnt bound is
substantially weakened or eliminated \cite{DPRV}. In \fig{bbneq} we
also give the expected $N_\nu$ value for different values of the
coupling $g$ between $\nu_\tau$'s and $J$'s, expressed in units of
$10^{-5}$. 
Comparing with the SM $g=0$ case one sees that for a fixed
$N_\nu^{max}$, a wide range of tau neutrino masses is allowed for
large enough values of $g$. No \nt masses below the LEP limit can be
ruled out, as long as $g$ exceeds a few times $10^{-4}$.
One can also see from the figure that {\sl $N_\nu$ can also be lowered
below the canonical SM value $N_\nu = 3$} due to the effect of the
heavy \nt annihilations to majorons.
These results may be re-expressed in the $m_{\nu_\tau}-g$ plane, as
shown in figure \ref{neffmg}.  
We note that the required values of $g(m_{\nu_\tau})$ fit well with
the theoretical expectations of many weak-scale majoron models.

The above discussion has been on the effect of \nt annihilations to
majorons in BBN.  In some weak-scale majoron models decays in \eq{NUJ}
may lead to short enough \nt lifetimes that they may also play an
important r\^ole in BBN \cite{bbnunstable}.

Before concluding the discussion on majorons, let me comment that the
majoron may be realized even in the context of models where B-L is a
gauge symmetry, such as left-right-symmetric models, by suitably
implementing a spontaneously broken global U(1) symmetry similar to
lepton number. It plays an interesting r\^ole in such models as it
allows the left-right scale to be relatively low~\cite{LRmajoron}.

\newpage
\section{Indications for Neutrino Mass}
\vskip .1cm

The most solid indications in favour of nonzero neutrino masses come
from underground experiments on solar and atmospheric neutrinos.  I
will provide a theorist's sketch of the present experimental
situation.

\subsection{Solar Neutrinos}
\vskip .1cm
 
The puzzle posed by the data collected by the Homestake, Kamiokande,
and the radiochemical Gallex and Sage experiments still defy an
explanation in terms of the Standard Model. The most recent data on
rates are summarized as: $2.56 \pm 0.23$ SNU (chlorine), $72.2 \pm
5.6$ SNU (Gallex and Sage gallium experiments sensitive to the $pp$
neutrinos), and $(2.44 \pm 0.10) \times 10^6 {\rm cm^{-2} s^{-1} }$
($^8$B flux from SuperKamiokande)~\cite{solarexp}.  This has been
re-confirmed by the 504 days data sample now collected by the
SuperKamiokande (SK) collaboration and reported at Neutrino 98~
\cite{sk504}.
\begin{figure}[t]
\centerline{\protect\hbox{
\psfig{file=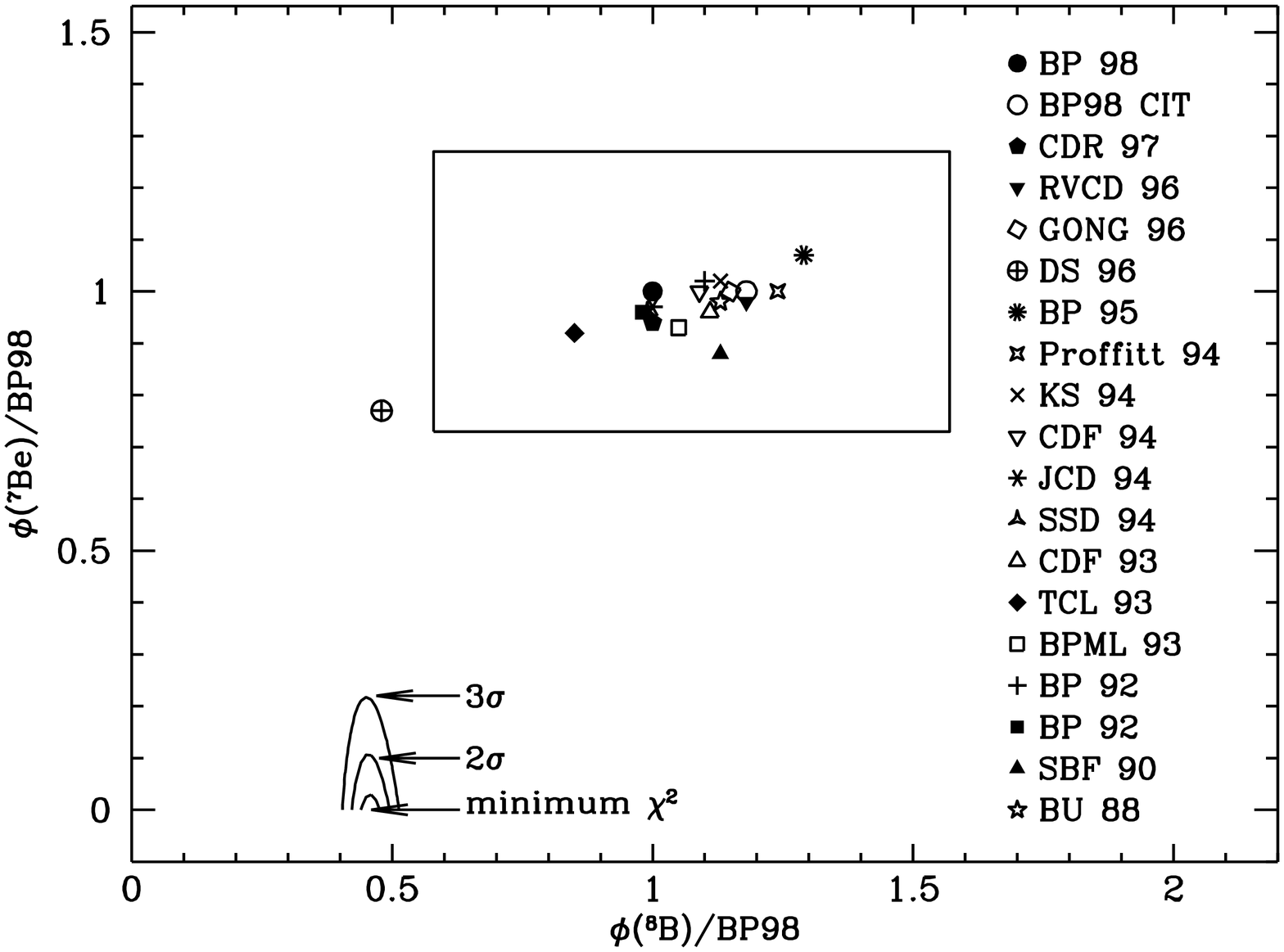,height=7cm,width=8cm}}}
\caption{Recent SSM predictions, from ref.~\protect\cite{Bahcall98}}
\label{78}
\end{figure}
In \fig{78} one can see the predictions of various standard solar
models in the plane defined by the $^7$Be and $^8$B neutrino fluxes,
normalized to the predictions of the BP98 solar model~\cite{BP98}.
Abbreviations such as BP95, identify different solar models, as given
in ref.~\cite{models}.  The rectangular error box gives the $3\sigma$
error range of the BP98 fluxes. The values of these fluxes indicated
by present data on neutrino event rates are also shown by the contours
in the figure. The best-fit $^7$Be neutrino flux is negative!
Possible non-standard astrophysical solutions are strongly constrained
by helioseismology studies \cite{Bahcall98} \cite{helio97}. Within the
standard solar model approach, the theoretical predictions clearly lie
far from the best-fit solution, and even far from the $3\sigma$
contour, leading us to conclude that new particle physics is the only
way to account for the data \cite{CF}.

The most likely possibility is to assume the existence of neutrino
conversions involving very small neutrino masses. The most attractive
theoretical schemes are the MSW effect \cite{MSW}, vacuum neutrino
oscillations or {\sl just-so solution} and, possibly, the Spin-Flavour
Precession mechanism proposed in ref.~\cite{SFP}, aided by the
Resonant enhancement due to matter effects in the Sun found in
ref.~\cite{RSFP}.  The resulting RSFP mechanism still provides a
viable solution to the solar neutrino problem \cite{akhmedov97}.

The recent SK data updates the 300 days situation we had before
Neutrino 98~\cite{sk300} without major surprises, except that the SK
collaboration has now given the first detailed report of the recoil
energy spectrum produced by solar neutrino
interactions~\cite{sk504}. The measured spectrum they reported at
Neutrino~98 shows more events at the highest bins than would have been
expected from the most popular neutrino oscillation parameters
discussed previously.  At first sight this might seem bad news for the
oscillation scenarios. However, Bahcall and Krastev have noted that if
the low energy cross section for ${\rm ^3He} ~+~ p ~\to ~{\rm ^4He}
~+~e^+ ~+~\nu_e $, the so-called $hep$ reaction, is $\gsim 20$ times
larger than the best (but uncertain) theoretical estimates, then this
reaction could significantly influence the electron energy spectrum
produced by solar neutrino interactions in the high recoil region.
This would hardly have any effect at lower energies.  They compare the
predicted energy spectra for different assumed $hep$ fluxes and
different neutrino oscillation scenarios with the one measured at
SuperKamiokande.
\begin{figure}
\centerline{\protect\hbox{
\psfig{file=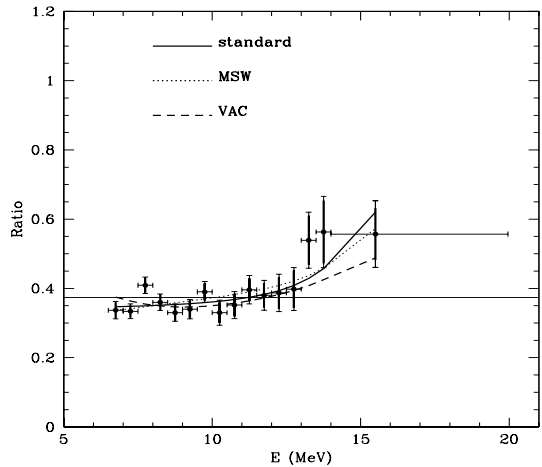,height=7cm,width=8cm}}}
\vglue -.5cm
\caption{Combined $^8$B plus $hep$ energy spectrum from
ref.~\protect\cite{bkhep}.  The total flux of $hep$ neutrinos was
varied so as to obtain the best-fit for each scenario.}
\label{spec500}
\end{figure}
Fig. \ref{spec500} shows the ratio of the measured \cite{sk504} to the
calculated number of events with electron recoil energy $E$.  The
crosses are the recent SK measurements~\cite{sk504}, while the
calculated curves are global fits to all of the data.  The horizontal
line at ${\rm Ratio} = 0.37$ represents the ratio of the total event
rate measured by SuperKamiokande to the predicted event
rate~\cite{BP98} with no oscillations and only $^8$B neutrinos. One
sees how the spectra with enhanced $hep$ contributions provide better
fits to the SK data, suggesting that these neutrinos may be playing a
r\^ole.

\begin{figure}
\centerline{
\protect\hbox{\psfig{file=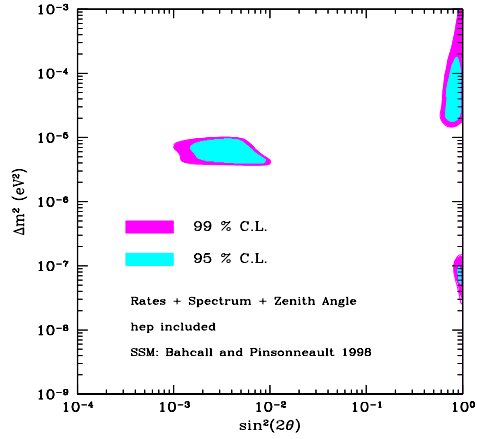,width=8cm,height=7cm}}}
\vglue -.5cm
\caption{Presently allowed MSW solar neutrino parameters for 2-flavour
active neutrino conversions with an enhanced $hep$  flux, from
ref.~\protect\cite{bkhep}}
\label{msw}
\end{figure}

One can determine the required solar neutrino parameters $\Delta m^2$
and $\sin^2 2\theta$ through a $\chi^2$ fit of the experimental data.
In \fig{msw} we show the allowed two-flavour regions obtained in an
updated MSW global fit analysis of the solar neutrino data for the
case of active neutrino conversions. The data include the chlorine,
Gallex, Sage~\cite{solarexp} and SK total event rates~\cite{sk504},
the SK energy spectrum~\cite{sk504}, as well as the SK day-night
asymmetry~\cite{sk504}, which would be expected in the MSW scheme due
to regeneration effects at the Earth. The data also includes the
recent SK~504~days sample. The analysis uses the BP98 model but with
an arbitrary $hep$ neutrino flux~\cite{bks98}.
One notices from the analysis that rate-independent observables, such
as the electron recoil energy spectrum and the day-night asymmetry
(zenith angle distribution), play an important r\^ole in ruling out
large regions of MSW parameters.

A theoretical issue which has raised some interest recently is the
study of the possible effect of random fluctuations in the solar
matter density \cite{BalantekinLoreti,noise,noise2}. The possible
existence of noise fluctuations at a few percent level is not excluded
by present helioseismology studies. 
\begin{figure}
\centerline{\protect\hbox{\psfig{file=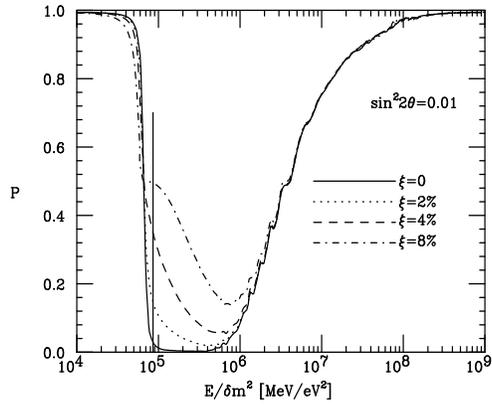,width=6.5cm,height=8cm,angle=90}}}
\vglue -.5cm
\caption{ Solar neutrino survival probability in the presence
of random density fluctuations, ref.~\protect\cite{noise}}
\label{Pnoise}
\end{figure}
In \fig{Pnoise} we show averaged solar neutrino survival probability
as a function of $E/\Delta m^2$, for $\sin^2 2\theta = 0.01$. This
figure was obtained via a numerical integration of the MSW evolution
equation in the presence of noise, using the density profile in the
Sun from BP95 in ref.~\cite{models}, and assuming that the correlation
length $L_0$ (which corresponds to the scale of the fluctuation) is
$L_0 = 0.1 \lambda_m$, where $\lambda_m$ is the neutrino oscillation
length in matter. An important assumption in the analysis is that $
l_{free} \ll L_0 \ll \lambda_m$, where $l_{free} \sim 10 $ cm is the
mean free path of the electrons in the solar medium. The fluctuations
may strongly affect the $^7$Be neutrino component of the solar
neutrino spectrum so that the Borexino experiment should provide an
ideal test, if sufficiently small errors can be achieved. The
potential of Borexino in probing the level of solar matter density
fluctuations provides an additional motivation for the experiment
\cite{borexino}. This is discussed in more detail in
ref. \cite{noise}.
\begin{figure}
\centerline{\protect\hbox{\psfig{file=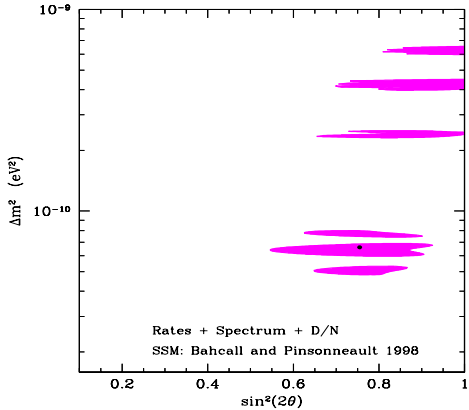,width=8cm,height=6.6cm}}}
\vglue -.5cm
\caption{Presently allowed vacuum oscillation
parameters, from ref.~\protect\cite{bks98}}
\label{vac98}
\end{figure}

The most popular alternative solution to the solar neutrino problem is
the {\sl vacuum oscillation solution} which clearly requires large
neutrino mixing and {\sl just-so} adjustment of the oscillation length
so as to coincide roughly with the Earth-Sun distance. This solution
fits with simplistic see-saw inspired-numerology and has attractive
features, as recently advocated in ref. \cite{glashow98}.
Fig. \ref{vac98} shows the regions of just-so oscillation parameters
obtained in a recent global fit of the data including the
504~days~SK~data sample, both the rates and the recoil energy
spectrum. Seasonal effects are expected in this scenario and could
potentially be used to further constrain the parameters, as described
in ref.~\cite{lisi}, and also to help discriminating it from the MSW
scenario.

\subsection{Atmospheric Neutrinos}
\vskip .1cm

Showers initiated when primary cosmic rays hit the Earth's atmosphere
originate secondary mesons, mostly pions and kaons, which decay
producing \ne's, \nm's as well as \bne's and \bnm's \cite{atmreview}.
There has been a long-standing discrepancy between the predicted and
measured \nm/\ne ratio of the atmospheric neutrino fluxes
\cite{atmexp}.  The anomaly was found both in water Cerenkov
experiments, such as Kamiokande, SuperKamiokande and IMB \cite{sk300},
as well as in the iron calorimeter Soudan2 experiment. Negative
experiments, such as Frejus and Nusex have much larger errors.

Although individual $\nu_\mu$ or $\nu_e$ fluxes are only known to
within $30\%$ accuracy, the $\nu_\mu$ $/\nu_e$ ratio is known to
$5\%$.  The most important feature of the atmospheric neutrino
535-day~ data~sample reported by the SK~collaboration at
Neutrino~98~\cite{superkatm98} is that it exhibits a {\sl
zenith-angle-dependent} deficit of muon neutrinos which is
inconsistent with expectations based on calculations of the
atmospheric neutrino fluxes. For recent analyses see
ref.~\cite{atm98,atmconcha,atmo98}. Experimental biases and
uncertainties in the prediction of neutrino fluxes and cross sections
are unable to explain the data.

\begin{figure}
\centerline{\protect\hbox{\epsfig{file=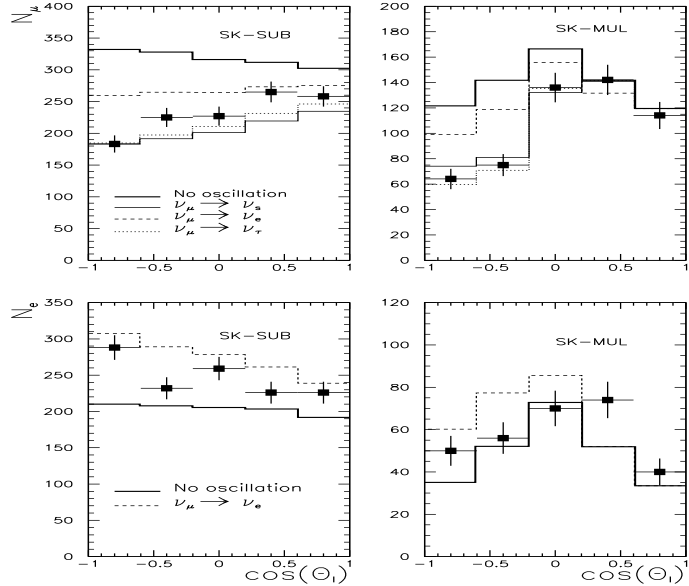,width=10cm,height=8.45cm}}}
\caption{Theoretically expected zenith angle distributions for SK
electron and muon-like sub-GeV and multi-GeV events in the SM
(no-oscillation) and for the best-fit points of the various
oscillation channels, from ref.~\protect\cite{atm98,atmconcha}. The
crosses correspond to the SK observations reported at Neutrino~98.}
\label{ang_mu}  
\end{figure}

In \fig{ang_mu} I show the measured zenith angle distribution of
electron-like and muon-like sub-GeV and multi-GeV events, as well as
the one predicted in the absence of oscillation. I also give the
expected distribution in various neutrino oscillation schemes.
The thick-solid histogram is the theoretically expected distribution
in the absence of oscillation, while the predictions for the best-fit
points of the various oscillation channels is indicated as follows:
for $\nu_\mu \to \nu_s$ (solid line), $\nu_\mu \to \nu_e$ (dashed
line) and $\nu_\mu \to \nu_\tau$ (dotted line).  The error displayed
in the experimental points is only statistical.

In the theoretical analysis we have used the latest improved
calculations of the atmospheric neutrino fluxes as a function of
zenith angle, including the muon polarization effect and took into
account a variable neutrino production point \cite{flux}.  Clearly the
data are not reproduced by the no-oscillation hypothesis, adding
substantially to our confidence that the atmospheric neutrino anomaly
is real.

\begin{figure}
\centerline{\protect\hbox{\epsfig{file=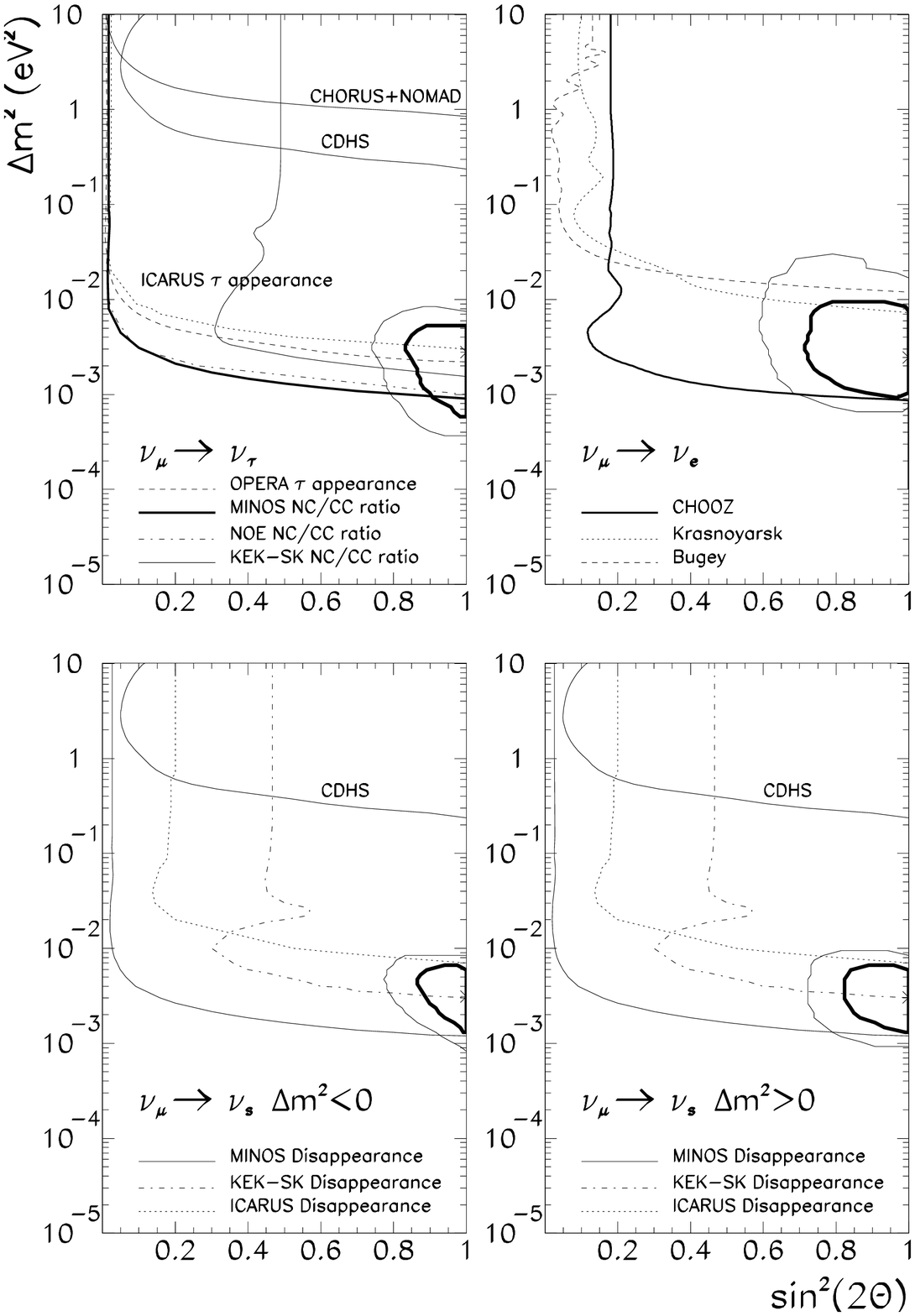,width=12.5cm,height=0.55\textheight}}}
\fcaption{Allowed atmospheric oscillation parameters for all
experiments including the SK data reported at Neutrino~98, combined at
90 (thick solid line) and 99 \% CL (thin solid line) for all possible
oscillation channels, from ref.~\protect\cite{atm98,atmconcha}.  In
each case the best-fit point is denoted by a star and always
corresponds to maximal mixing, a feature which is well-reproduced by
the theoretical predictions of the models proposed in
ref.~\protect\cite{ptv92,pv93}. The sensitivity of the present
accelerator and reactor experiments as well as the expectations of
upcoming long-baseline experiments is also displayed.}
\label{mutausk4} 
\end{figure}

In \fig{mutausk4} I show the allowed neutrino oscillation parameters
obtained in a recent global fit of the sub-GeV and multi-GeV
(vertex-contained) atmospheric neutrino data~\cite{atm98,atmconcha}
including the recent data reported at Neutrino~98, as well as all
other experiments combined at 90 (thick solid line) and 99 \% CL (thin
solid line) for each oscillation channel considered.
The two lower panels \fig{mutausk4} differ in the sign of the $\Delta
m^2$ which was assumed in the analysis of the matter effects in the
Earth for the $\nu_\mu \to \nu_s$ oscillations. We found that $\nu_\mu
\to \nu_\tau$ oscillations give a slightly better fit than $\nu_\mu
\to \nu_s$ oscillations.  At present the atmospheric neutrino data
cannot distinguish between the \nm to \nt and \nm to \ns channels.  
It is well-know that the neutral-to-charged current ratios are
important observables in neutrino oscillation phenomenology, which are
especially sensitive to the existence of singlet neutrinos, light or
heavy \cite{2227}.
The atmospheric neutrinos produce isolated neutral pions
($\pi^0$-events) mainly in neutral current interactions.
One may therefore study the ratios of $\pi^0$-events and the events
induced mainly by the charged currents, as recently advocated in
ref.~\cite{vissani}. This has the virtue of minimizing uncertainties
related to the original atmospheric neutrino fluxes.
In fact the SK~collaboration has already tried to do this by
estimating the double ratio of $\pi^0$ over e-like events in their
sample~\cite{superkatm98} and found $R = 0.93 \pm 0.07 \pm 0.19$.
This is consistent both with \nm to \nt or \nm to \ns channels, with a
slight preference for the former. The situation should improve in the
future.

We also display in \fig{mutausk4} the sensitivity of present
accelerator and reactor experiments, as well as that expected at
future long-baseline (LBL) experiments in each channel.  The first
point to note is that the Chooz reactor \cite{Chooz} data already
excludes the region indicated for the $\nu_{\mu} \to \nu_e$ channel
when all experiments are combined at 90\% CL.

From the upper-left panel in \fig{mutausk4} one sees that the regions
of $\nu_\mu \to \nu_\tau$ oscillation parameters obtained from the
atmospheric neutrino data analysis cannot be fully tested by the LBL
experiments, as presently designed.  One might expect that, due to the
upward shift of the $\Delta m^2$ indicated by the fit for the sterile
case (due to the effects of matter in the Earth) it would be possible
to completely cover the corresponding region of oscillation
parameters. Although this is the case for the MINOS disappearance
test, in general most of the LBL experiments can not completely probe
the region of oscillation parameters allowed by the $\nu_\mu \to
\nu_s$ atmospheric neutrino analysis.  This is so irrespective of the
sign of $\Delta m^2$ assumed.  For a discussion of the various
potential tests that can be performed at the future LBL experiments in
order to unravel the presence of oscillations into sterile channels
see ref.~\cite{atmconcha}.

\subsection{LSND, Dark Matter \& Pulsars}

\vskip .1cm
{\sl LSND}
\vskip .1cm

\begin{figure}
\centerline{\protect\hbox{\epsfig{file=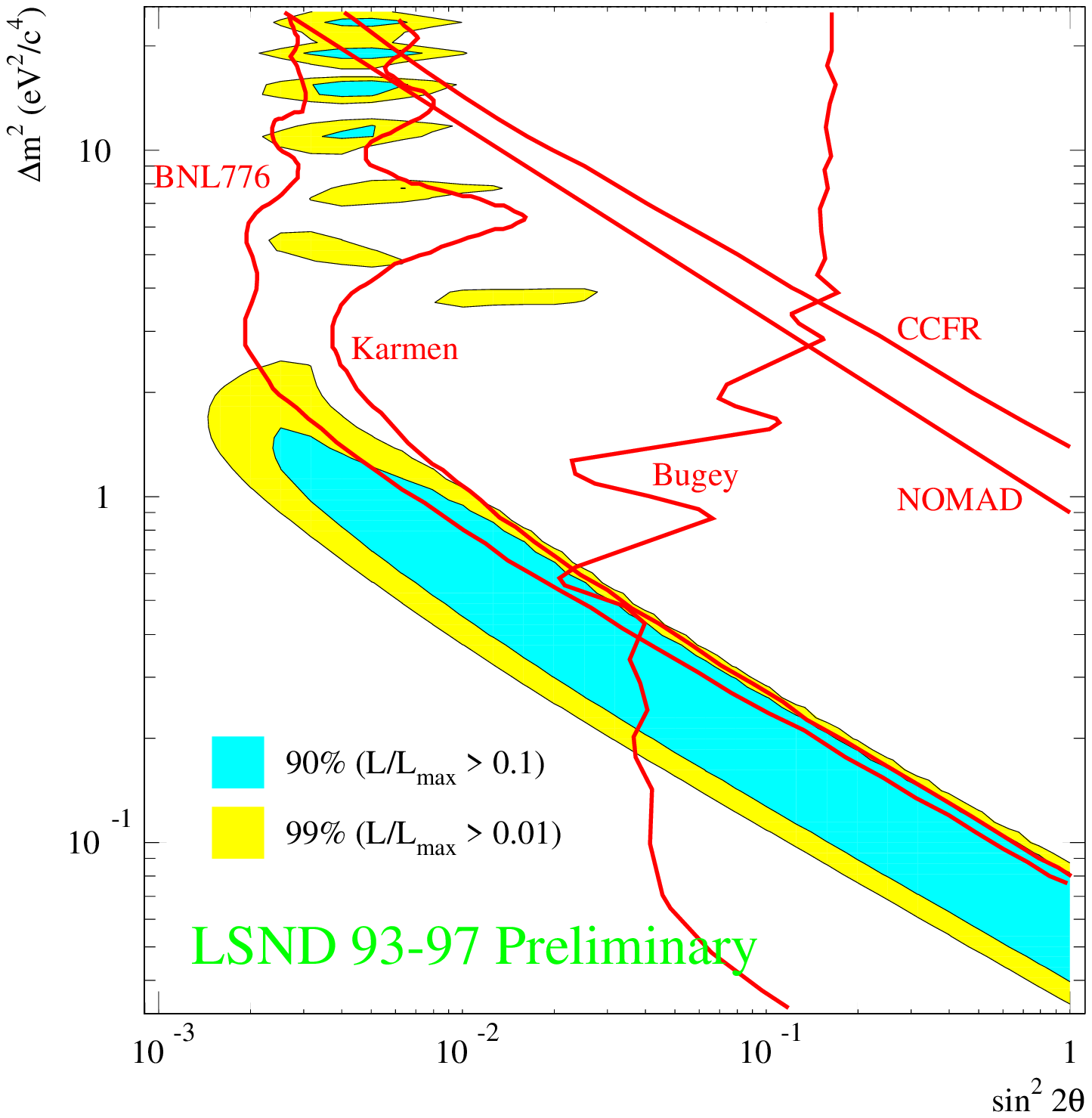,width=8cm,height=8cm}}}
\fcaption{Allowed LSND oscillation parameters versus competing
experiments~\protect\cite{louis} }
\label{darlsnd} 
\vglue -.5cm
\end{figure}

\begin{figure}
\centerline{\protect\hbox{\epsfig{file=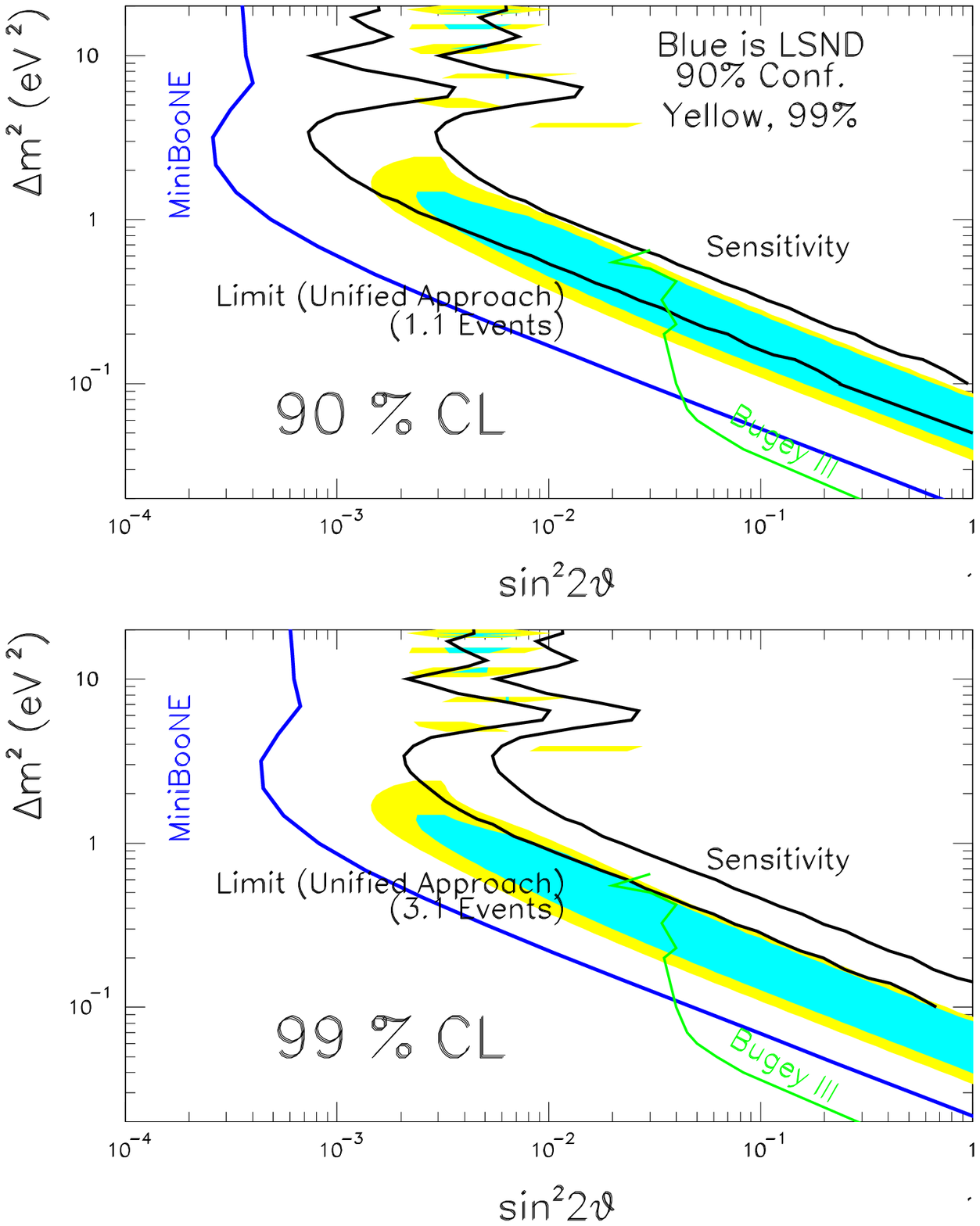,width=8cm,height=10cm}}}
\fcaption{Expected sensitivity of the proposed MiniBooNE
experiment~\protect\cite{louis} }
\label{miniboone} 
\end{figure}

A search for $\bar\nu_{\mu}\to \bar\nu_{e}$ oscillations has been
conducted at the Los Alamos Meson Physics Facility by using
$\bar\nu_\mu$ from $\mu^+$ decay at rest \cite{LSND}. The
$\bar\nu_e$'s are detected via the reaction $\bar\nu_e\,p \to
e^{+}\,n$, correlated with a $\gamma$ from $np \to d \gamma$
($2.2\,{\rm MeV}$). The use of tight cuts to identify $e^+$ events
with correlated $\gamma$ rays yields 22 events with $e^+$ energy
between 36 and $60\,{\rm MeV}$ and only $4.6 \pm 0.6$ background
events. A fit to the $e^+$ events between 20 and $60\,{\rm MeV}$
yields a total excess of $51.8^{+18.7}_{-16.9} \pm 8.0$ events. If
attributed to $\bar \nu_\mu \to \bar \nu_e$ oscillations, this
corresponds to an oscillation probability of ($0.31^{+0.11}_{-0.10}
\pm 0.05$)\% and leads to the oscillation parameters shown in
\fig{darlsnd}. The shaded regions are the favoured likelihood regions
given in ref.~\cite{LSND}. The curves show the 90~\% and 99~\%
likelihood allowed ranges from LSND, and compares them to limits from
BNL776, KARMEN1, Bugey, CCFR, and NOMAD.
A search for \nm $\to$ \ne oscillations has also been conducted by the
LSND collaboration.  Using \nm from $\pi^+$ decay in flight, the \ne
appearance is detected via the charged-current reaction
$C(\ne,e^-)X$. Two independent analyses are consistent with the above
signature, after taking into account the events expected from the \ne
contamination in the beam and the beam-off background.  If interpreted
as an oscillation signal, the observed oscillation probability of $2.6
\pm 1.0 \pm 0.5 \times 10^{-3}$ is consistent with the \bnm $\to$ \bne
oscillation evidence described above. Fig.~\ref{miniboone} compares the
LSND region with the expected sensitivity from MiniBooNE, which was
recently approved to run at Fermilab~\cite{louis}.
A possible confirmation of the LSND anomaly would be a discovery of
far-reaching implications.

\vskip .2cm
{\sl Dark Matter}
\vskip .1cm

The research on the nature of the cosmological dark matter and the
origin of galaxies and large scale structure in the Universe within
the standard theoretical framework of gravitational collapse of
fluctuations as the origin of structure in the expanding universe has
undergone tremendous progress recently. Indeed the observations of
cosmic background temperature anisotropies on large scales performed
by the COBE satellite \cite{cobe} combined with cluster-cluster
correlation data e.g. from IRAS~\cite{iras} can not be reconciled with
the simplest cold dark matter (CDM) model.
Barring a non-zero cosmological constant and high value of the Hubble
parameter ($h \gsim 0.7$) the simplest model that have a chance to
work is Cold + Hot Dark Matter (MDM, for mixed dark matter), if the
Hubble parameter and age parameter allow for an $\Omega=1$ cosmology
\cite{cobe2}, suggested by inflation. Electron-volt mass neutrinos are
the most well-motivated HDM candidate. This mass scale is similar to
that indicated by the hints reported by the LSND experiment
\cite{LSND}.

However it is too early to be confident on the MDM scenario, and one
should for the moment keep an open mind. For example, I note that an
MeV range (unstable) tau neutrino is an interesting possibility to
consider from the point of view of dark matter.  If such neutrino
decays before the matter dominance epoch, its decay products would add
energy to the radiation, thereby delaying the time at which the matter
and radiation contributions to the energy density of the universe
become equal. Such delay would allow one to reduce the density
fluctuations on the smaller scales purely within the standard cold
dark matter scenario \cite{ma1}.

Future sky maps of the cosmic microwave background radiation (CMBR)
with high precision at the upcoming MAP and PLANCK missions should
bring more light into the nature of the dark matter and the possible
r\^ole of neutrinos \cite{Raffelt}.

\vskip .2cm
{\sl Pulsars}
\vskip .1cm

One of the most challenging problems in modern astrophysics is to find
a consistent explanation for the high velocity of pulsars.
Observations \cite{veloc} show that these velocities range from zero
up to 900 km/s with a mean value of $450 \pm 50$ km/s.  An attractive
possibility is that pulsar motion arises from an asymmetric neutrino
emission during the supernova explosion. In fact, neutrinos carry more
than $99 \%$ of the new-born proto-neutron star's gravitational
binding energy so that even a $1 \%$ asymmetry in the neutrino
emission could generate the observed pulsar velocities.  One possible
explanation to this puzzle may reside in the interplay between the
parity non-conservation present in weak interactions and the strong
magnetic fields which are expected during a SN explosion.  
Possible realizations of this idea in the framework of the Standard
Model (SM) have been proposed \cite{Chugai,others} However, it has
recently been noted~\cite{vilenkin98} that no asymmetry in neutrino
emission can be generated in thermal equilibrium, even in the presence
of parity violation. This suggests that alternative mechanism is at
work.
Several neutrino conversion mechanisms in matter have been invoked as
a possible engine for powering pulsar motion.  They all share in
common the feature that neutrino propagation properties are affected
by the {\sl polarization} \cite{NSSV} of the SN medium which is
provided by the strong magnetic fields $10^{15}$ Gauss present during
a SN explosion. This would give rise to some angular dependence of the
matter-induced neutrino potentials leading to a deformation of the
"neutrino-sphere" for, say, tau neutrinos and hence to an anisotropic
neutrino emission.  As a consequence, in the presence of non-vanishing
$\nu_\tau$ mass and mixing the resonance sphere for the
$\nu_e-\nu_\tau$ conversions is distorted.  If the resonance surface
lies between the $\nu_\tau$ and $\nu_e$ neutrino spheres, such a
distortion would induce a temperature anisotropy in the flux of the
escaping tau-neutrinos produced by the conversions, hence a recoil
kick of the proto-neutron star.
This mechanism was realized in ref.~\cite{KusSeg96} invoking MSW
conversions \cite{MSW} with \mnt $\gsim$ 100 eV or so, assuming a
negligible $\nu_e$ mass. This is necessary in order for the resonance
surface to be located between the two neutrino-spheres.  It should be
noted, however, that such requirement is at odds with cosmological
bounds on neutrinos masses unless the $\tau$-neutrino is unstable.
On the other hand in ref.~\cite{ALS} a realization was proposed in the
resonant spin-flavour precession scheme (RSFP) \cite{RSFP}.  Here the
magnetic field not only affects the medium properties, but also
induces the spin-flavour precession through its coupling to the
neutrino transition magnetic moment \cite{SFP}. 
Perhaps the simplest suggestion was proposed in ref.~\cite{pulsars}
where the required pulsar velocities would arise from anisotropic
neutrino emission induced by resonant conversions of massless
neutrinos (hence no magnetic moment) \cite{massless0}. This mechanism
arises in the model described in \eq{MATmu} and has been shown to be
of potential relevance for SN physics \cite{massless}. 

Very recently, however, Raffelt and Janka~\cite{pulsars2} have claimed
that the asymmetric neutrino emission effect was vastly
overestimated, because the variation of the temperature over the
deformed neutrino-sphere is not an adequate measure for the anisotropy
of the neutrino emission. This would invalidate the oscillation
mechanisms, leaving the pulsar velocity problem without any known
viable solution. The only potential way out of their criticism would
invoke conversions into sterile neutrinos, since the conversions would
take place deeper in the star. However, it is too early to tell
whether or not it works \cite{nuno98}.

\section{Reconciling the neutrino puzzles}
\vskip .1cm
 
It is easy to accommodate the solar and atmospheric neutrino data by
themselves in a general gauge theory of neutrino mass, since it lacks
predictivity. One could even have a situation where three-neutrino
mixing could be {\sl bi-maximal}, i.e. maximal in both the atmospheric
as well as solar neutrino transitions, if the solution chosen by
nature is {\sl just-so} \cite{glashow98}.
The challenge to reconcile these two requirements arise mainly if one
wishes to do that in a predictive quark-lepton {\sl unification}
scheme that relates lepton and quark mixing angles. This especially so
since the latter are small, in contrast to the lepton mixing indicated
by the SK~atmospheric data.
The story gets more complicated if one wishes to account also for the
LSND anomaly and for the hot dark matter. There has been a lot of
effort to solve the bigger puzzle posed by the inclusion of any of
these additional hints~\cite{ptv92,pv93,cm93}.  As we have seen the
atmospheric neutrino data requires $\Delta m^2_{atm}$ which is much
larger than the scale $\Delta {m^2}_\odot$ which is indicated by the
solar neutrino data, either in the context of the MSW mechanism or the
just-so solution. These two experiments fix two different scales for
neutrino mass differences, so that with just the three known neutrinos
and without discarding any experimental data, there is no room to
include the LSND scale indicated in \fig{darlsnd}, nor the HDM scale
which is roughly similar
\footnote{I will ignore the pulsar velocity problem since there is
no clear working-model at the moment.}.

Reconciling the neutrino puzzles may be attempted within the {\sl
unification approach} or the {\sl weak-scale approach} to the theory
of neutrino mass. I will concentrate mostly on the latter, because it
is an interesting and simpler alternative to the former.

\subsection{Almost Degenerate Neutrinos}
\vskip .1cm

The only possibility to fit solar, atmospheric and HDM scales in a
world with just the three known neutrinos is if all of them have
nearly the same mass \cite{cm93}, of about $\sim$ 1.5 eV or so in
order to provide the right amount of HDM \cite{cobe2} (all three
active neutrinos contribute to HDM). There is no room in this case to
accommodate the LSND anomaly. This can be arranged in the unification
approach discussed in sec. 2 using the $M_L$ term present in general
in seesaw models. With this in mind one can construct, e.g. unified
\10 seesaw models where all neutrinos lie at the above HDM mass scale
($\sim$ 1.5 eV), due to a suitable horizontal symmetry, while the
parameters $\Delta {m^2}_\odot$ \& $\Delta {m^2}_{atm}$ appear as
symmetry breaking effects. An interesting fact is that the ratio
$\Delta {m^2}_\odot \:/\:\Delta {m^2}_{atm}$ appears as
${m_c}^2/{m_t}^2$~\cite{DEG}.

\subsection{Four-Neutrino Models}
\vskip .1cm

The simplest way to open the possibility of incorporating the LSND
scale is to invoke a sterile neutrino, i.e. one whose interaction with
standard model particles (such as the $W$ and the $Z$) is much weaker
than the SM weak interaction.  It must come in as an \21 singlet
ensuring that it does not affect the invisible Z decay width,
well-measured at LEP. The sterile neutrino \ns must also be light
enough in order to participate in the oscillations involving the three
active neutrinos. The theoretical challenges we have are:
\bi 
\item
to understand why the sterile neutrino is so light (it is clear that
if a sterile neutrino is introduced into the SM, the \21 gauge
symmetry allows it to have a bare mass, which could be large)
\item
to account for the maximal neutrino mixing indicated by the
atmospheric data  
\item
to account for the three scales $\Delta m^2_{atm}$, $\Delta
{m^2}_\odot$ and $\Delta m^2_{LSND/HDM}$ from first principles  
\ei
With this in mind we have formulated the simplest and first
schemes~\cite{ptv92,pv93} which provide an answer to the above points.
I will denote them, $(e\tau)(\mu~s)$~\cite{ptv92} and $(es)(\mu\tau)$
~\cite{pv93}, respectively. One should realize that a given
phenomenological scheme (mainly determined by the structure of the
leptonic charged current) may be realized in more than one theoretical
model. For example, an alternative to the model in ~\cite{pv93} was
suggested in ref.~\cite{cm93}. There have been many attempts to
reproduce the above phenomenological scenarios from different
theoretical assumptions, as has been discussed here
\cite{ptvlate,smir,4nutalks}.
 
These two basic schemes are characterized by a very symmetric mass
spectrum in which there are two ultra-light neutrinos at the solar
neutrino scale and two maximally mixed almost degenerate eV-mass
neutrinos (LSND/HDM scale), split by the atmospheric neutrino
scale~\cite{ptv92,pv93}. The HDM problem requires 
the heaviest neutrinos at about 2 eV mass \cite{pvhdm}.
These scales are generated radiatively due to the additional Higgs
bosons which are postulated, as follows: $\Delta m^2_{LSND/HDM}$
arises at one-loop, while $\Delta m^2_{atm}$ and $\Delta {m^2}_\odot$
are two-loop effects.  Since this proposal pre-dated the LSND results,
it naturally focussed on accounting for the HDM problem, rather than
LSND. However, it has been realized that the LSND oscillation effects
may be accounted for in its framework. These are the simplest theories
based only on weak-scale physics, in which one {\sl explains} the
lightness of the sterile neutrino, the large lepton mixing required by
the atmospheric neutrino data, as well as the generation of the mass
splittings responsible for solar and atmospheric neutrino
conversions. These follow naturally from the underlying
lepton-number-like symmetry and its breaking~\cite{ptv92,pv93}.

\begin{figure}[t]
\centerline{\protect\hbox{\psfig{file=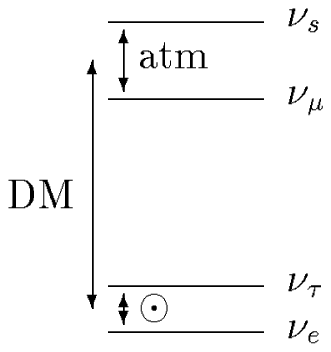,width=6cm,height=4cm}}}
\caption{$(e\tau)(\mu~s)$ scheme: \ne- \nt conversions explain the
solar neutrino data and \nm- \ns oscillations account for the
atmospheric deficit, ref.~\protect\cite{ptv92}.}
\label{ptv}
\vglue -.2cm
\end{figure}

\begin{figure}
\centerline{\protect\hbox{\psfig{file=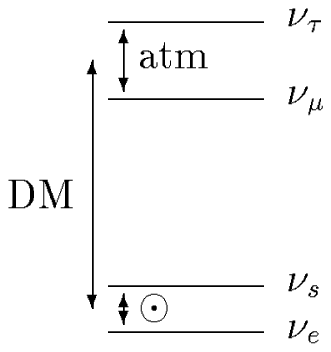,width=6cm,height=4cm}}}
\caption{$(es)(\mu\tau)$ scheme: \ne- \ns conversions explain the
solar neutrino data and \nm- \nt oscillations account for the
atmospheric deficit, ref.~\protect\cite{pv93}.}
\label{pv}
\vglue -.2cm
\end{figure}

These models are minimal in the sense that they add a single \21
singlet lepton to the SM.  Before breaking the symmetry the heaviest
neutrinos are exactly degenerate, while the other two which will be
responsible for the explanation of the solar neutrino problem are
still massless \cite{OLDsterilemodel}. After the global U(1) lepton
symmetry breaks the massive ones split and the light ones get mass.
The models differ according to whether the \ns lies at the dark matter
scale or at the solar neutrino scale. In the $(e\tau)(\mu~s)$ scheme
the \ns lies at the LSND/HDM scale, as illustrated in \fig{ptv}
while in the alternative $(es)(\mu\tau)$ model, \ns is at the solar
\neu scale as shown in \fig{pv} \cite{pv93} \cite{ptvlate}.
In the $(e\tau)(\mu~s)$ case the atmospheric \neu puzzle is explained
by \nm to \ns oscillations, while in $(es)(\mu\tau)$ it is explained
by \nm to \nt oscillations. Correspondingly, the deficit of solar
\neus is explained in the first case by \ne to \nt conversions, while
in the second the relevant channel is \ne to $\nu_s$. The two models
are therefore clearly inequivalent. In both cases it is possible to
fit all present observations together.

I now turn to the consistency of the models with BBN.  The presence of
additional weakly interacting light particles, such as our light
sterile neutrino \ns, is constrained by BBN since the \ns would enter
into equilibrium with the active neutrinos in the early Universe (and
therefore would contribute to $N_\nu^{max}$) via neutrino oscillations
\cite{bbnsterile}, unless
%
$\Delta m^2 sin^42\theta \lsim 3\times 10^{-6}~~eV^2$
%
where $\Delta m^2$ denotes the mass-square difference of the active
and sterile species and $\theta$ is the vacuum mixing angle.  However,
systematical uncertainties in the derivation of BBN bounds still
caution us not to take them too literally. For example, it has been
argued in~\cite{sarkar} that present observations of primordial Helium
and deuterium abundances can allow up to $N_\nu = 4.5$ neutrino
species if the baryon to photon ratio is small. Adopting this as a
limit, clearly both models described above are consistent. Should the
BBN constraints get tighter, e.g. $N_\nu^{max} < 3.5$ they could rule
out the $(e\tau)(\mu~s)$ model, and leave out only the competing
scheme as a viable alternative. For recent work on this see
ref.~\cite{volkasterilebbn}.

The two models would be distinguishable both from the analysis of
future solar as well as atmospheric neutrino data. For example they
may be tested in the SNO experiment~\cite{SNO} once they measure the
solar neutrino flux ($\Phi^{NC}_{\nu}$) in their neutral current data
and compare it with the corresponding charged current value
($\Phi^{CC}_{\nu}$). If the solar neutrinos convert to active
neutrinos, as in the $(e\tau)(\mu~s)$ model, then one expects
$\Phi^{CC}_{\nu}/\Phi^{NC}_{\nu} \simeq .5$, whereas in the
$(es)(\mu\tau)$ scheme (\ne conversion to \ns), the above ratio would
be nearly $ \simeq 1$.  Looking at pion production via the neutral
current reaction $\nu_{\tau} + N \to \nu_{\tau} +\pi^0 +N$ in
atmospheric data might also help in distinguishing between these two
possibilities~\cite{vissani}, since this reaction is absent in the
case of sterile neutrinos, but would exist in the $(es)(\mu\tau)$
scheme.

If light sterile neutrinos indeed exist, as suggested by the current
solar and atmospheric neutrino data, together with the LSND
experiment, one can show that in some four-neutrino scenarios,
neutrinos would contribute to a cosmic hot dark matter component and
to an increased radiation content at the epoch of matter-radiation
equality. These effects leave their imprint in sky maps of the cosmic
microwave background radiation (CMBR) and may thus be detectable with
the precision measurements of the upcoming MAP and PLANCK missions
as noted recently in ref.~\cite{Raffelt}.

\subsection{MeV Tau Neutrino}
\vskip .1cm

In ref.~\cite{JV95} a model was presented where an unstable MeV
Majorana tau neutrino naturally reconciles the cosmological
observations of large and small-scale density fluctuations with the
cold dark matter picture (CDM). The model assumes the spontaneous
violation of a global lepton number symmetry at the weak scale.  The
breaking of this symmetry generates the cosmologically required decay
of the \nt with lifetime $\tau_{\nu_\tau} \sim 10^2 - 10^4$ sec, as
well as the masses and oscillations of the three light \neus \ne, \nm
and $\nu_s$. One can also verify that the BBN constraints can be
satisfied.  The cosmological attractiveness of this scheme should
encourage one to check whether one can indeed account for the present
solar and atmospheric data through oscillations among the three light
neutrinos, after taking into account the recent SKdata.

\section{In conclusion}
\vskip .1cm

A major news has been the re-confirmation of an angle-dependent
atmospheric neutrino deficit by the SK collaboration, providing a
strong evidence for neutrino masses, similar to that offered by the
solar neutrino data. Unfortunately future LBL experiments do not all
probe the full region indicated by the atmospheric data.
If the LSND result stands the test of time, this would be a puzzling
indication for the existence of a light sterile neutrino. {\sl Who
ordered it?}
The two most attractive schemes to reconcile these observations invoke
either \ne- \nt conversions to explain the solar data, with \nm- \ns
oscillations accounting for the atmospheric deficit, or the other way
around. These two basic schemes have distinct implications at future
solar \& atmospheric neutrino experiments. SNO and SuperKamiokande
have the potential to distinguish them due to their neutral current
sensitivity.

{\sl How about heavy neutrinos?}  Although cosmological bounds are a
fundamental tool to restrict neutrino masses, in many theories heavy
neutrinos will either decay or annihilate very fast, thereby loosening
or evading the cosmological bounds. From this point of view, {\sl
neutrinos can have any mass presently allowed by laboratory
experiments}, and it is therefore important to search for
manifestations of heavy neutrinos at the laboratory.

Last but not least, though most of the recent excitement comes from
underground experiments, one should note that models of neutrino mass
may lead to a plethora of new signatures which may be accessible also
at accelerators, thus illustrating the complementarity between the two
approaches in unravelling the properties of neutrinos and probing for
signals beyond the SM.

\vskip .1cm

I am grateful to Bernd Kniehl and Georg Raffelt for the kind
hospitality at the Ringberg castle. My thanks to John Bahcall, Plamen
Krastev and Bill Louis, for making their postcript figures available
to me, and to Thomas Janka and Eligio Lisi for correspondence. I thank
all my collaborators, especially Hiroshi Nunokawa for going over the
first draft of this manuscript critically.  This work was supported by
DGICYT grant PB95-1077 and by the EEC under the TMR contract
ERBFMRX-CT96-0090.

\end{document}